\documentstyle[aps,epsf]{revtex}

\begin{document}
\draft
\title{A Mesoscopic Approach to the ``Negative'' Viscosity Effect in Ferrofluids}
\author{A. P\'erez-Madrid, T. Alarc\'on, J.M.G. Vilar, and J. M. Rub\'{\i
}}
\address{Departament de F\'{\i}sica Fonamental\\
Facultat de F\'{\i}sica\\
Universitat de Barcelona\\
Diagonal 647, 08028 Barcelona, Spain\\
}
\maketitle

\begin{abstract}
We present a mesoscopic approach to analyze the dynamics of a single magnetic dipole under the influence of an oscillating magnetic field, based on the formulation of a Fokker-Planck equation. The dissipated power and the viscosity of a suspension of such magnetic dipoles are calculated from non-equilibrium thermodynamics of magnetized systems. By means of
 this method we have found a non-monotonous behaviour of the viscosity as 
a function of the frequency of the field which has been referred to as the ``negative'' viscosity effect. Moreover, we have shown that the viscosity depends on the vorticity field thus exhibiting non-Newtonian behaviour. Our 
analysis is complemented with numerical simulations which reproduce the behaviour of the viscosity we have  found and extend the scope of our analytical approach to higher values of the magnetic field.
\end{abstract}


\section{introduction}

It is well-known that the dynamics of a suspension of dipolar particles is strongly influenced by the presence of an external field. Concerning the viscosity of the suspension, its behaviour as a function of the magnetic field is monotonous in the case of a constant field, reaching a saturation limit where the magnetic moments are oriented along the field. In contrast, when the field oscillates that behaviour is considerably modified 
to the extent that the contribution of the oscillating field to the effective viscosity of the suspension may become negative for frequencies of 
the field larger than the local vorticity. This phenomenon observed experimentally in \cite{kn:bacri} and reported in refs. \cite{kn:rosensweig} and  \cite{kn:shliomis} has been referred to as the ``negative'' viscosity effect in ferrofluids. It reveals the presence of two regimes, one essentially 
dissipative in which the variation of the viscosity is positive and other
in which the energy of the oscillating field is practically transformed 
into kinetic energy of the particle. It is in this last regime where
the variations of the viscosity become negative.

 Our purpose in this paper is to present an explanation of that effect based on a Fokker-Planck dynamics describing 
the time evolution of the probability density for the orientation of the 
dipolar particle. The starting point is the formulation of that equation 
and its perturbative solution. In this way we compute the different components of the susceptibility. The dissipated power and the viscosity follow 
from an analysis based on non-equilibrium thermodynamics.

The general methodology we introduce accounts for the results of ref. 
\cite{kn:shliomis}, valid in the limit of small field, and agree with the experimental data of ref. \cite{kn:bacri}. Our theoretical results are also compared with numerical simulations we have performed, indicating that the qualitative behavior of the system is satisfactorily reproduced even for a moderate intensity of the
 oscillating magnetic field.

The paper is organized as follows. Using linear response theory, we compute in Section II the generalized
susceptibility associated with the orientation vector of the magnetic
dipoles in the ferrofluid, when this is under the influence of an
oscillating magnetic field. Section III is devoted to obtain the
contributions of the oscillating field to the dissipation of energy and to
the effective viscosity. In Section IV we report results on the viscosity obtained from numerical simulations, whereas in Section V we present our main conclusions.

\section{Response of a magnetic dipole to an oscillating field}

We consider a dilute colloidal suspension of ferromagnetic dipolar spherical
particles \cite{kn:brenner2}, with magnetic moment, $\vec{m}= m_{s}\hat{
\vec{R}}$, where $\hat{\vec{R}}$ is an unit vector accounting for the
orientation of the dipole. Each dipole is under the influence of a vortex flow with vorticity $\vec{\Omega}= 2\omega _{0}\hat{\vec{z}}$, with 
$\hat{%
\vec{z}}$ being the unit vector along the $z$-axis, and of an oscillating field       $\vec{H}= He^{-i\omega t}\hat{\vec{x}}$, with $\hat{\vec{x}}$ being the 
unit
vector along the $x$-axis. For $t\gg \tau _{r}$, the motion of the particle is overdamped with $\tau _{r}= I/\xi _{r}$ being the inertial time scale. Here $I$ is the moment of inertia, $\xi _{r}= 8\pi \eta_0 a^{3}$ is the
rotational friction coefficient, with $\eta_0 $ the solvent viscosity, and $a$ 
the radius of the particle. This time scale defines a  cut-off frequency
$%
\omega _{r}= \tau _{r}^{-1}$, such that the condition for overdamped motion
is equivalent to $\omega \ll \omega _{r}$. In this case, the balance of the magnetic and hydrodynamic torques acting on each particle

\begin{equation}  \label{eq:r1}
\vec{m}\times\vec{H} + \xi_r\left(\frac{1}{2}\vec{\Omega} - \vec{\Omega
}%
_p\right) = 0 ,
\end{equation}

\noindent together with the rigid rotor evolution equation

\begin{equation}  \label{eq:r2}
\frac{d \hat{\vec{R}}}{d t} = \vec{\Omega}_p\times\hat{\vec{R}}
\end{equation}

\noindent lead to the dynamic equation for $\hat{\vec{R}}$ 

\begin{equation}  \label{eq:r3}
\frac{d \hat{\vec{R}}}{d t} = \left\{\omega_0\hat{\vec{z}} \right.+ 
\lambda
(t)\left.\left (\hat{\vec{R}}\times\hat{\vec{x}}\right )\right\}\times\hat
{%
\vec{R}}.
\end{equation}

\noindent Here $\lambda (t)\equiv (m_{s}H/\xi _{r})e^{-i\omega t}$
,
with $\vec{\Omega}_{p}$ being the angular velocity of the particle.

With allowance for Brownian motion, the stochastic dynamics corresponding
 to
eq. (\ref{eq:r3}) is given by the Fokker-Planck equation, involving the
probability density $\Psi(\hat{\vec{R}},t)$

\begin{equation}  \label{eq:a1}
\partial_{t}\Psi(\hat{\vec{R}},t)\,=\,({\cal L}_{0}+\lambda(t){\cal L}
_{1})\Psi(\hat{\vec{R}},t),
\end{equation}

\noindent where ${\cal L}_{0}$ and ${\cal L}_{1}$ are operators defined by

\begin{eqnarray}  \label{eq:a2}
{\cal L}_{0}&=&-\omega_{0}\,\hat{\vec{z}}\cdot{\cal \vec{R}}+D_{r}{\cal
 \vec{%
R}}^{2},  \nonumber \\
&& \\
{\cal L}_{1}&=&2\,\hat{\vec{R}}\cdot\hat{\vec{x}}- 
\,(%
\hat{\vec{R}}\times\hat{\vec{x}})\cdot{\cal \vec{R}},  \nonumber
\end{eqnarray}

\noindent with $D_{r}=k_{B}T/\xi _{r}$ being the rotational diffusion
coefficient, and ${\cal \vec{R}}=\hat{\vec{R}}\times \partial /\partial
 \hat{%
\vec{R}}\;$ the rotational operator. Notice that the first and second terms
 on
the right hand side of eq. (\ref{eq:a2})$_{1}$, correspond to convective
and diffusive term, respectively. Moreover eq. (\ref{eq:a1}) which, 
according to eq. (\ref{eq:r3}), rules the Brownian dynamics in the case of
overdamped motion, is valid in the diffusion regime. This regime is also
characterized by the condition $t\gg \tau _{r}$, or equivalently $\omega 
\ll
\omega _{r}$, which implicitly involves the white noise assumption.

To solve the Fokker-Planck equation (\ref{eq:a1}) we will assume that $%
\lambda _{0}\equiv \vert \lambda(t)\vert$ constitutes a small parameter such that this equation
can be solved perturbatively. Thus up to first order in $\lambda $, the
solution of the Fokker-Planck equation (\ref{eq:a1}) is

\begin{equation}  \label{eq:a5}
\Psi(\hat{\vec{R}},t)\,=\,e^{(t-t_{0}){\cal L}_{0}}\Psi^{0}(t_{0})+%
\int_{t_{0}}^{t}\,dt^{\prime}\,\lambda(t^{\prime})e^{(t-t^{\prime}){\cal 
L}%
_{0}}{\cal L}_{1}\Psi^{0}(t^{\prime})
\end{equation}

\noindent Here $\Psi^{0}(t^{\prime}) = e^{(t^{\prime}-t_0){\cal L}%
_{0}}\Psi^{0}(t=t_{0})$ is the zero order solution at time $t^{\prime}$
, and

\begin{equation}  \label{eq:a3}
\Psi^{0}(\hat{\vec{R}},t =t_0)\,=\,\delta(\hat{\vec{R}}-\hat{\vec{R}}_0) ,
\end{equation}

\noindent with $\hat{\vec{R}}_0$ being an arbitrary initial orientation. 
As follows from eq. (\ref{eq:a2})$_1$, the unperturbed operator ${\cal L
}
_{0}$ is composed of the operators ${\cal R}_{z}$ and ${\cal R}^{2}$, which
are proportional to the orbital angular momentum operators of quantum
mechanics $L_{z}$ and $L^{2}$, respectively, and, therefore, their
eigenfunctions are the spherical harmonics\cite{kn:sakurai}

\begin{eqnarray}  \label{eq:a6}
{\cal R}_{z}Y_{l\, m}(\hat{\vec{R}})&=& imY_{l\, m}(\hat{\vec{R}}), \nonumber \\
{\cal R}^{2}Y_{l\, m}(\hat{\vec{R}})&=& -l(l+1)Y_{l\, m}(\hat{\vec{R}}) .
\end{eqnarray}

Given that we know how $\vec{{\cal R}}$ acts on the spherical harmonics, 
it
is convenient to expand the initial condition in series of these functions,
since the spherical harmonics constitute a complete set of functions which
are a basis in the Hilbert space of the integrable functions over the unit
sphere\cite{kn:hilbert}

\begin{equation}  \label{eq:a8}
\Psi^{0}(\hat{\vec{R}},t_0)\, =\,\delta(\hat{\vec{R}}-\hat{\vec{R}}_0)\,
=\,\sum_{l=0}^{\infty}\,\sum_{m=-l}^{l}\, Y_{l\, m}^{*}(\hat{\vec{R
}}%
_0)Y_{l\, m}(\hat{\vec{R}} )\;\; .
\end{equation}

\noindent Using this expansion in eq. (\ref{eq:a5}), for the first order
correction to the probability density, $\Delta\Psi\equiv\Psi-\Psi^{0}$, we obtain

\begin{equation}  \label{eq:a9}
\Delta\Psi(\hat{\vec{R}},t)\,
=\,\sum_{l=0}^{\infty}\;\sum_{m=-l}^{l}\;\int_{t_{0}}^{t}\,
dt^{\prime}\lambda(t^{\prime})Y_{l\, m}^{*}(\hat{\vec{R}}_0)e^{(t-t^{\prime})%
{\cal L}_0}{\cal L}_1 e^{(t^{\prime}-t_0){\cal L}_{0}}Y_{l\, m}(\hat{\vec
{R}}%
) .
\end{equation}

\noindent Notice that the integral of $\Delta\Psi(\hat{\vec{R}},t)$ over 
the
entire solid angle is zero, in agreement with the fact that the unperturbed
solution $\Psi^0(\hat{\vec{R}},t)$ is normalized.

Since, we are interested in the asymptotic behavior we will set $
t_{0}\rightarrow -\infty$. In this limit, eq. (\ref{eq:a5}) becomes

\begin{equation}  \label{eq:m1}
\Psi (\hat{\vec{R}} ,t) = \frac{1}{4\pi}\left\{ 1 + \int_{-\infty}^{t}\right.\;
dt^{\prime}\,\lambda (t^{\prime })\,e^{(t-t^{\prime }){\cal L}_{0}}2
\left.\right.
\left. \hat{\vec{R}}\cdot\hat{\vec{x}}\right\},
\end{equation}

\noindent where now

\begin{equation}  \label{eq:m2}
\Delta \Psi (\hat{\vec{R}},t)=\frac{1}{4\pi }\int_{-\infty}^{t} \;dt^{\prime}\,\lambda (t^{\prime })\,e^{(t-t^{\prime }){\cal L}_{0}}\,2
\hat{
\vec{R}}\cdot \hat{\vec{x}},
\end{equation}
\noindent and 

\begin{equation}  \label{eq:r30}
\Psi ^{0}(\hat{\vec{R}},t)=\frac{1}{4\pi}
\end{equation}

\noindent is the uniform distribution function in the unit sphere.

>From eq. (\ref{eq:m2}) the contribution of the AC field to the mean value of
the orientation vector $\hat{\vec{R}}$ can be obtained as

\begin{equation}  \label{eq:a10}
\hat{\vec{R}}(t)\,=\,\int\, d\hat{\vec{R}}\, \hat{\vec{R}} \,\Delta\Psi
\,
=\, \frac{1}{4\pi}\int_{-\infty}^{t}\;dt^{\prime}\lambda(t^{ \prime})\int\, d%
\hat{\vec{R}}\, \hat{\vec{R}}\, e^{(t-t^{\prime}){\cal L} _0}\, 2
\hat{\vec{R}}\cdot\hat{\vec{x}}.
\end{equation}

\noindent This equation can be written in the more compact form

\begin{equation}
\hat{R}_{i}(t)\,=\,\int_{-\infty }^{t}\,dt^{\prime }\lambda (t^{\prime
})\chi _{i}(t-t^{\prime }),  \label{eq:a11}
\end{equation}

\noindent where the response function\cite{kn:resibois} has been
defined as

\begin{equation}  \label{eq:a12}
\chi _{i}(\tau )\,=\,\frac{1}{4\pi }\int \,d\hat{\vec{R}}\,\hat{R}%
_{i}e^{\tau {\cal L}_{0}}\,2\hat{\vec{R}}\cdot \hat{\vec{x}},
\end{equation}

\noindent for $\tau >0$.

By causality, we can write $t\rightarrow \infty $ in the upper limit of the
integral in eq. (\ref{eq:a11}); hence, this equation becomes

\begin{equation}
\hat{R}_{i}(t)\,=\,\chi _{i}(\omega )\lambda (t).  \label{eq:a13}
\end{equation}

\noindent where $\chi _{i}(\omega )$ is the generalized susceptibility,
 which is the Fourier transform of $\chi _{i}(\tau )\,$ 

\begin{equation}  \label{eq:a14}
\chi _{i}(\omega )\,=\,\frac{1}{4\pi }\int_{-\infty }^{\infty }d\tau
e^{i\omega \tau }\int d\hat{\vec{R}}\hat{R}_{i}e^{\tau {\cal L}_{0}}2 \hat{\vec{R}}\cdot \hat{\vec{x}}
\end{equation}

\noindent From this equation we obtain the components of the susceptibility

\begin{eqnarray}  \label{eq:f1}
\chi _{x}(\omega )\,=\, &\frac{1}{3}\left\{ \left[ \frac{2D_{r}}{%
4D_{r}^{2}+(\omega -\omega _{0})^{2}}-i\frac{(\omega _{0}-\omega )}{%
4D_{r}^{2}+(\omega -\omega _{0})^{2}}\right] +\right. &  \nonumber \\ \vspace{.25cm}
&\left. \left[ \frac{2D_{r}}{4D_{r}^{2}+(\omega +\omega _{0})^{2}}+i\frac
{%
(\omega _{0}+\omega )}{4D_{r}^{2}+(\omega +\omega _{0})^{2}}\right] \right\}
&
\end{eqnarray}

\begin{eqnarray}  \label{eq:f2}
\chi_y(\omega)\, =\,&\frac{1}{3}\left\{\left[\frac{(\omega_0
-\omega)}{4D_r^2 + (\omega-\omega_0)^2}+i\frac{2D_r}{4D_r^2 +
(\omega-\omega_0)^2}\right]+\right .&  \nonumber \\ \vspace{.25cm}
&\left .\left[\frac{(\omega_0 +\omega)}{4D_r^2 + (\omega+\omega_0)^2}-i\frac{%
2D_r}{4D_r^2 + (\omega+\omega_0)^2}\right]\right\}&
\end{eqnarray}

\begin{equation}  \label{eq:f3}
\chi_z(\omega)\, =\, 0
\end{equation}

\noindent The quantities $\chi_x$, and $\chi_y$, have poles at $\omega = \pm\omega_0 \pm 2
D_ri$%
. The inverse of the imaginary part of these poles, $%
(2D_r)^{-1}$, defines the Brownian relaxation time.

\section{Non-equilibrium thermodynamics of the relaxation process}

Our purpose in this section is to compute the energy dissipated during the relaxation process of the magnetization and the rotational viscosity of the suspension. The starting point is the entropy production corresponding to the relaxation of the magnetization, as given in ref. \cite{kn:mazur}  

\begin{equation}  \label{eq:m3}
\sigma = -\frac{1}{T}\frac{d\vec{M^{\prime}}}{dt}\cdot (\vec{H}_{eq}^{\prime}-\vec{H}^{\prime})\; .
\end{equation}

\noindent Here $\vec{M}$ is the magnetization of the suspension and $\vec
{H}$ the magnetic field. Moreover, $\vec{H}_{eq}$ is the magnetic field related to the instantaneous value of $\vec{M}$. The primes indicate that 
the corresponding quantities have been computed in the frame of reference
 rotating with the fluid.

The entropy production can alternatively be written in terms of the corresponding quantities in the laboratory frame. Using the relation between the temporal derivatives of the magnetization in both frames
\begin{equation}
\frac{d\vec{M^{\prime}}}{dt} = \frac{d\vec{M}}{dt} - \frac{1}{2} \vec{\Omega}\times\vec{M}\; .
\label{eq:g1}
\end{equation}

\noindent  from eq. (\ref{eq:m3})  one obtains

\begin{eqnarray}\label{eq:f4}
\sigma &=& -\frac{1}{T}\left\{\frac{d\vec{M}}{dt} - \frac{1}{2} \vec{\Omega}\times\vec{M}\right\}\cdot (\vec{H}_{eq}-\vec{H}) = \\ \nonumber
 \vspace{.25cm}
& &-\frac{1}{T}\frac{d\vec{M}}{dt}\cdot (\vec{H}_{eq}-\vec{H}) + \frac{1}{2T} \vec{\Omega}\times\vec{M}\cdot (\vec{H}_{eq}-\vec{H}) \;
  \end{eqnarray}

\noindent where now the different quantities refer to the laboratory system. Notice that in our case $\vec{M}=c\; m_s\hat{\vec{R}}$, where $c$ is the concentration of particles. Moreover, eq. (\ref{eq:f4}) written in this way ensures the frame material invariance of the entropy production.

The linear law inferred from this expression coincides with the relaxation equation postulated by Shliomis\cite{kn:shliomis}, provided we identify 
the phenomenological coefficient with the inverse of the Brownian relaxation  time.

The form of eq. (\ref{eq:f4}) suggests the following decomposition 

\begin{equation}\label{eq:j4}
\sigma = \sigma_D + \sigma_V \; .
 \end{equation}
\noindent The first contribution 

\begin{equation}\label{eq:j1}
\sigma_D = -\frac{1}{T}\frac{d\vec{M}}{dt}\cdot (\vec{H}_{eq}-\vec{H}) 
= -\frac{1}{T}\frac{d\vec{M}}{dt}\cdot \vec{H}
\end{equation}
\noindent accounts for the entropy production coming from Debije relaxation, whereas the second

\begin{equation}\label{eq:j2}
\sigma_V = \frac{1}{2} (\vec{\Omega}\times\vec{M})\cdot (\vec{H}_{eq}-\vec{H}) = -\frac{1}{T} \vec{\Omega}\cdot(\vec{M}\times\vec{H})
\end{equation}

\noindent is related to the viscous dissipation. To obtain these expressions use has been made of the relation $\vec{M} = \kappa\vec{H}_{eq}$, with $\kappa$ being the static susceptibility, and of the fact that $M$ remains constant which implies $(d\vec{M}/dt)\cdot\vec{H}_{eq} = 0$, and $(\vec{\Omega}\times\vec{
M})\cdot\vec{H}_{eq} = 0$.

The power dissipated in a period of the field follows from the entropy production we have computed. For the different contributions one has

\begin{equation}\label{eq:j5}
P_{\alpha}(\omega) = \frac{\omega}{2\pi}\int_0^{2\pi/\omega} T\sigma_{\alpha} dt\; ,\;\; (\alpha = D,V)\; .
\end{equation}

\noindent For the Debije contribution we obtain

\begin{eqnarray}\label{eq:d1}
P_D(\omega)\,&=&\; 6\eta_0 \phi\omega\; Im\chi_x (\omega)\; \overline{%
\lambda(t)^{2}} = \\ \nonumber \vspace{.25cm}
& & \eta_0 \phi \omega \lambda _{0}^{2}\left[ \frac{%
\omega +\omega _{0}}{4D_{r}^{2}+(\omega +\omega _{0})^{2}}+\frac{\omega
-\omega _{0}}{4D_{r}^{2}+(\omega -\omega _{0})^{2}}\right]  \label{p6}
\end{eqnarray}

\noindent where we have used eqs. (\ref{eq:a13}) and (\ref{eq:f1}), whith 
 $\phi = (4\pi/3)a^3\, c$ being the volume fraction of particles. Similarly, by using eqs. (\ref{eq:a13}) and (\ref{eq:f2}) the viscous part yields

\begin{eqnarray}\label{eq:j6}
P_V(\omega) &=& 6 \eta_0\phi \omega_{0}\;Re\chi _{y}(\omega )\;\overline{
\lambda (t)^{2}} = \\ \nonumber \vspace{.25cm}
& &\eta_0 \phi \omega _{0}\lambda _{0}^{2}\left[ \frac{\omega_{0}
+\omega }{4D_{r}^{2}+(\omega_{0} +\omega )^{2}}+\frac{\omega _{0}-\omega }{4D_{r}^{2}+(\omega_{0} -\omega )^{2}}\right] \; 
\end{eqnarray}

\noindent This last contribution introduces the rotational viscosity defined through the relation

\begin{equation}\label{eq:j7}
P_V(\omega) = \eta_{r}(2\omega_0)^2\; .
\end{equation}

\noindent Combining eqs. (\ref{eq:j6}) and (\ref{eq:j7}) one infers the value of the rotational viscosity

\begin{equation}\label{eq:j8}
\eta_r = \frac{1}{4} \eta_0 \phi \omega _{0}^{-1}\lambda _{0}^{2}\left[
\frac{\omega_{0}
+\omega }{4D_{r}^{2}+(\omega_{0} +\omega )^{2}}+\frac{\omega _{0}-\omega}
{4D_{r}^{2}+(\omega_{0} -\omega )^{2}}\right] \; 
\end{equation}

\noindent which represents the contribution of the rotational degrees of freedom of the dipoles to the effective viscosity of the suspension given by

\begin{equation}\label{eq:s2}
\eta_{eff} = \eta + \eta_r
\end{equation}

\noindent where $\eta$ is given by Einstein law $\eta = \eta_0(1 + 5/2\phi)$, 
 \cite{kn:agusti}. In view of the former results, the total dissipation $P(\omega) = P_D(\omega) + P_V(\omega)$ is

\begin{equation}\label{eq:p7}
P(\omega) = \eta_0 \phi \lambda _{0}^{2}\left[ \frac{%
(\omega_{0} +\omega )^2}{4D_{r}^{2}+(\omega_{0} +\omega )^{2}}+\frac{(\omega_{0}
-\omega )^2}{4D_{r}^{2}+(\omega_{0} -\omega )^{2}}\right]\; .
\end{equation}

\noindent Notice that although the contributions to the dissipation  $P_D
(\omega)$ and $P_V(\omega)$ may achieve negative values, the total dissipation  remains always positive, in accordance with the second law.

The linear law derived from the entropy production (24) is valid for situations closed to equilibrium, when the distribution function is not too different from the equilibrium distribution. For larger deviations, this approach and equivalently the relaxation 
equation of Shliomis are no longer valid. One then should employ the Fokker Planck description we have introduced in section II. In this way, results 
for higher values of the field could also be obtained by means of the for
malism we have developed.

\section{Numerical simulations}

In order to check the validity of our results and explore the behaviour of the system for higher values of the 
oscillating field we have performed numerical simulations  by using a 
standard  second-order Runge-Kutta  method for stochastic differential  equations \cite{sde1,sde2}. To this purpose we have considered the Langevin equation corresponding to eq. (\ref{eq:a1}):

\begin{equation}  \label{eq:l1}
\frac{d \hat{\vec{R}}}{d t} = \left\{\omega_0\hat{\vec{z}} \right.+\lambda
(t)\left.\left (\hat{\vec{R}}\times\hat{\vec{x}}\right)+ \vec{\xi}(t)\right\} 
\times\hat{\vec{R}},
\end{equation}

\noindent where $\vec{\xi}(t)$ is Gaussian  white  noise with zero mean and correlation \mbox{$\left<\xi_i(t)\xi_j(t+\tau)\right>=2D_r\delta_{ij}\delta(\tau)$}.

>From the previous equation one can easily compute the mean angular velocity
of the particles by averaging over several realizations of the noise. This quantity then gives the rotational viscosity 

\begin{equation}\label{eq:s3}
\eta_r={3 \over 2}\eta_0\phi\left(1-{\Omega_p \over \omega_0}\right) \;\,
\end{equation}

\noindent To obtain this expression we will first rewrite the entropy production in terms of $\vec{P}^a$, the axial vector related to the antisymmetric part of the pressure tensor

\begin{equation}\label{eq:s1}
\sigma = \frac{1}{T}\vec{P}^a\cdot \vec{\Omega}\;,
\end{equation}
\noindent from which one derives the phenomenological law 
\begin{equation}\label{eq:g2}
\vec{P}^a = \eta_r \vec{\Omega}\; .
\end{equation}
\noindent The balance of torque densities,  which is achieved in the asymptotic regime \cite{kn:agusti}, 
\begin{equation}\label{eq:g3}
2\vec{P}^a = \vec{M}\times\vec{H} = 6\eta_0\phi (\frac{1}{2}\vec{\Omega} - \vec{\Omega}_p)
\end{equation}
\noindent together with eqs. (\ref{eq:g2}), and (\ref{eq:g3}) then leads to expression  (\ref{eq:s3}).

\begin{figure}[th]
\centerline{
\epsfxsize= 20.0 cm 
\epsffile{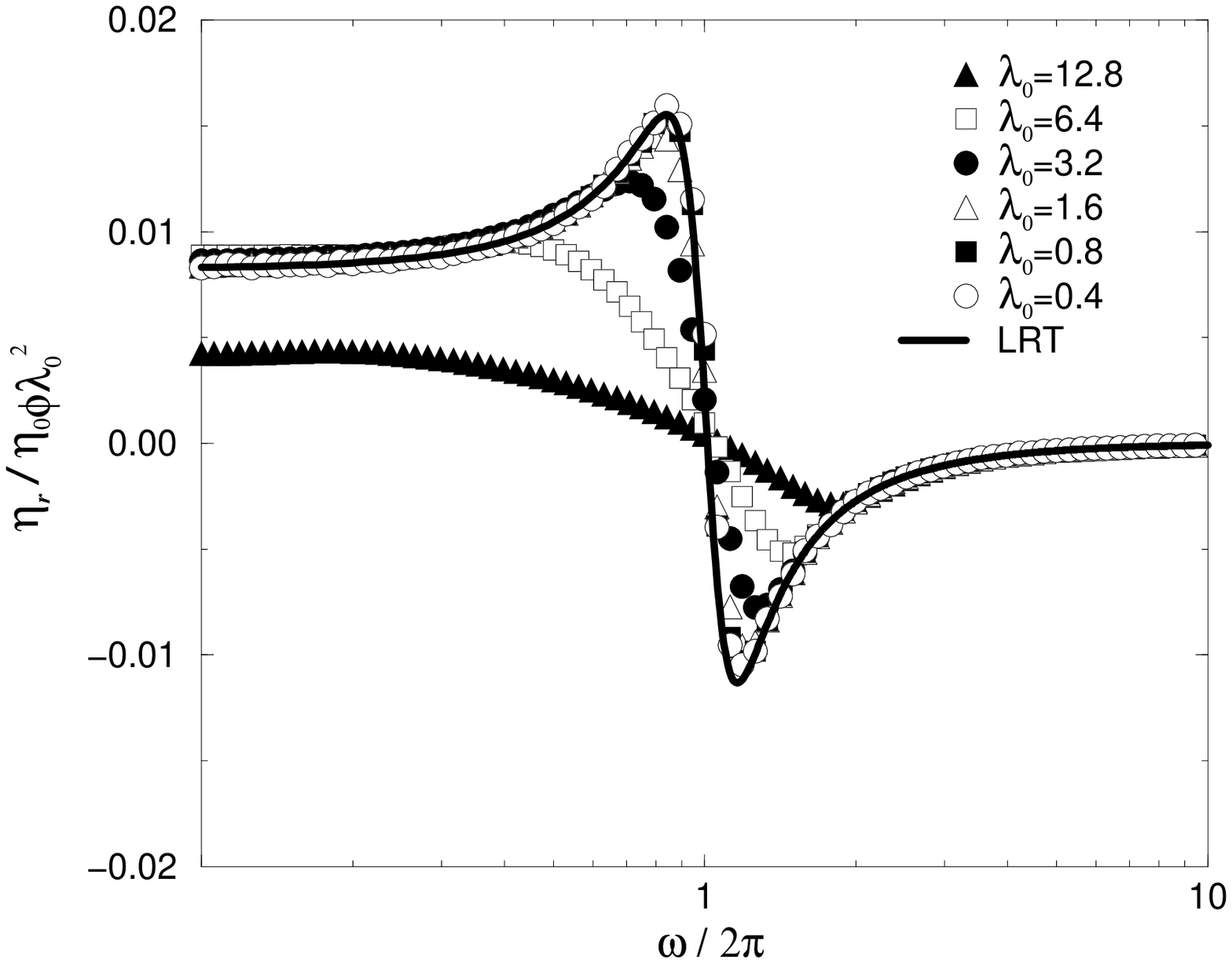}}
\caption[f]{\label{fig1}
Normalized rotational viscosity, $\eta_r/\eta_0\phi\lambda^2_0$, as a function of the frequency of the applied field for $\lambda_0=\{12.8, \;6.4,
 \;3.2, \;1.6, \;0.8, \;0.4\}$ and linear response theory (LRT).
The values of the remaining parameters are $D_r=0.5$ and $\omega_0=2\pi$.}
\end{figure}

\begin{figure}[th]
\centerline{
\epsfxsize= 20.0 cm 
\epsffile{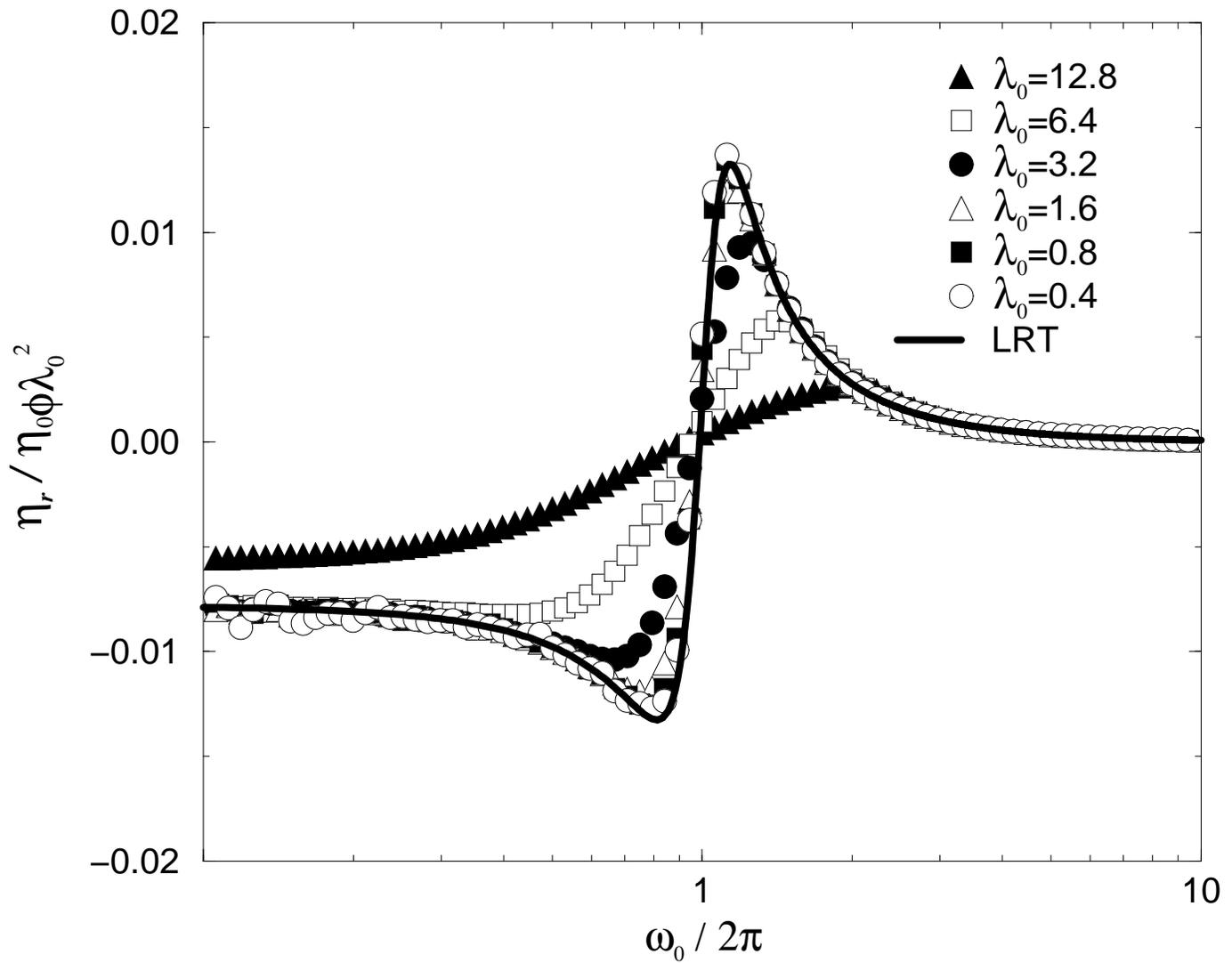}}
\caption[f]{\label{fig2}
Normalized rotational viscosity, $\eta_r/\eta_0\phi\lambda^2_0$, as a function of the angular velocity of the fluid for $\lambda_0=\{12.8, \;6.4, 
\;3.2, \;1.6, \;0.8, \;0.4\}$ and linear response theory (LRT).
The values of the remaining parameters are $D_r=0.5$ and $\omega=2\pi
$.}
\end{figure}

In Figs. (\ref{fig1}) and (\ref{fig2}) we  show the results obtained  for  the normalized rotational viscosity, $\eta_r/\eta_0\phi\lambda
^2_0$, as a function of the frequency of the applied field and as a function of the angular velocity of the fluid, respectively, for different intensities of the oscillating field. These figures clearly illustrate how for low intensities of the oscillating field the results obtained through numerical simulations can accurately be reproduced by linear response theory, as well as for higher values, 
linear response theory is still qualitatively correct. The
 crossover point from positive to negative values of the rotational viscosity
 seems not to depend significantly on the amplitude of the oscillating field. It
 is interesting to realize that, at high frequencies all the simulation curves
 match the linear response theory curve. Thus, in this frequency regime linear response theory provides an accurate description of the phenomenon. Fig. (\ref{fig2}) also makes the non-Newtonian character of the 
fluid manifest. Notice that these effects are more pronounced for angular
 velocities of the fluid near the angular frequency of the applied field,
i.e., for high and low values of the vorticity the rotational viscosity goes to
zero and  a constant negative value, respectively, whereas for
intermediate values it depends on the particular value of the vorticity. From both figures, one concludes that the vorticity plays just the opposite role to the frequency of the field.

\section{Conclusions}

In this paper we have presented a mesoscopic approach to explain the \mbox{``negative''} viscosity effect ocurring in suspensions of magnetic particles. The dynamics of the magnetic moment in the oscillating field is described by means of a Fokker-Planck equation which can be solved perturbatively. This equation gives rise to a hierarchy of equations\cite{kn:carmen} for the different moments, describing the relaxation of the magnetic 
moment. The first equation of the hierarchy, when linearized in the field
, agrees with the phenomenological equation obtained from non-equilibrium
 thermodynamics and the corresponding one postulated by Shliomis.

Following this procedure, we have been able to compute the dissipated power and the viscosity which is a non-monotonous function of the frequency 
of the field. Up to first order in the field our results agree with the corresponding ones of ref. \cite{kn:shliomis} based on the phenomenological equation proposed by Shliomis.

The phenomenological approach, dealing with the phenomenological law derived from the entropy production, is strictely  valid at small fields and is no longer correct 
when larger deviations from the equilibrium distribution occur. If we are
 interested in the response of the system to larger values of the field the correct approach is the one based upon the Fokker-Planck equation we have proposed in section II. This approach involves the evolution equation for higher order moments of the probability distribution in the hierarchy. After introducing a decoupling approximation, the solution for the first moment can be used to compute the energy dissipation that allows one to define an effective viscosity.

In order to go beyond the linear regime we have performed numerical simulations. This numerical analysis enables us to discern about the
validity of the linear response theory treatment. Our conclusion is that
qualitatively linear response theory may provide a reasonably explanation of the phenomenon.

\acknowledgments

This work has been
supported by DGICYT of the Spanish Government under grant PB95-0881, and
also by the INCO-COPERNICUS program of the European Commission under
contract IC15-CT96-0719. T. Alarc\'{o}n wishes to thank DGICYT
of the Spanish Government for financial support.

\pagebreak

\vfill
\pagebreak

\end{document}